\documentclass[aps,prd,reprint,a4paper,showpacs,nofootinbib,superscriptaddress]{revtex4-2}

\usepackage{bbm}
\usepackage{amsmath}
\usepackage{graphicx}
\usepackage{amssymb}
\usepackage{subfigure}
\usepackage{amssymb}
\usepackage{hyperref}
\usepackage{epstopdf}

\usepackage[utf8]{inputenc}
\hypersetup{
    colorlinks=true,
    linkcolor=red,
    citecolor=blue,
}
\usepackage{color}
\usepackage[T1]{fontenc}
\usepackage{txfonts}
\usepackage{mathrsfs}
\usepackage{latexsym}
\usepackage{epsfig}
\usepackage{epstopdf}
\usepackage{epstopdf}
\usepackage{graphicx}
\usepackage{amssymb}
\usepackage{amsmath}
\usepackage{dcolumn}
\usepackage{bm}
\usepackage{color}
\usepackage{comment}
\usepackage{xcolor}
\usepackage{dsfont}

\usepackage{graphicx}
\usepackage{dcolumn}
\usepackage{bm}
\usepackage{color}
\usepackage{enumitem}
\usepackage{amsmath}
\usepackage{amssymb}

\newcommand{\beq}{\begin{equation}}
\newcommand{\eeq}{\end{equation}}
\newcommand{\bea}{\begin{eqnarray}}
\newcommand{\eea}{\end{eqnarray}}

\usepackage{bbm}
\usepackage{amsfonts}
\usepackage{mathrsfs}
\usepackage{latexsym}
\usepackage{epsfig}
\usepackage{epstopdf}
\usepackage{epstopdf}
\usepackage{graphicx}
\usepackage{amssymb}
\usepackage{amsmath}
\usepackage{dcolumn}
\usepackage{bm}
\usepackage{color}
\usepackage{comment}
\usepackage{xcolor}

\usepackage{graphicx}
\usepackage{dcolumn}
\usepackage{bm}
\usepackage{color}
\usepackage{enumitem}
\usepackage{amsmath}
\usepackage{amssymb}

\usepackage{bbm}
\usepackage{amsfonts}
\usepackage{mathrsfs}
\usepackage{latexsym}
\usepackage{epsfig}
\usepackage{epstopdf}
\usepackage{epstopdf}
\usepackage{graphicx}
\usepackage{amssymb}
\usepackage{amsmath}
\usepackage{dcolumn}
\usepackage{bm}
\usepackage{color}
\usepackage{comment}
\usepackage{xcolor}

\newcommand{\Order}{\mathcal{O}}
\newcommand{\ee}{\mathrm{e}}
\newcommand{\mpl}{M_{\rm Pl}}
\newcommand{\dd}{\mathrm{d}}
\newcommand{\as}{A_s}
\newcommand{\nbd}{\mathrm{BD}}
\newcommand{\ex}{\mathrm{ex}}
\newcommand{\zetaP}{\mathcal{P}_{\zeta}}
\newcommand{\tensorP}{\mathcal{P}_{t}}

\begin{document}

\title{Long Inflation Screens Euclidean-Wormhole Initial States}

\author{Imtiaz Khan}
\email{ikhanphys1993@gmail.com}
\affiliation{Department of Physics, Zhejiang Normal University, Jinhua, Zhejiang 321004, China}
\affiliation{Research Center of Astrophysics and Cosmology, Khazar University, Baku, AZ1096, 41 Mehseti Street, Azerbaijan}

\author{Pirzada}
\email{pirzada@itp.ac.cn}
\affiliation{CAS Key Laboratory of Theoretical Physics, Institute of Theoretical Physics, Chinese Academy of Sciences, Beijing 100190, China}
\affiliation{School of Physical Sciences, University of Chinese Academy of Sciences, No. 19A Yuquan Road, Beijing 100049, China}

\author{G. Mustafa}
\email{gmustafa3828@gmail.com}
\affiliation{Department of Physics, Zhejiang Normal University, Jinhua, Zhejiang 321004, China}

\author{Farruh~Atamurotov}
\email{atamurotov@yahoo.com}
\affiliation{University of Tashkent for Applied Sciences, Str. Gavhar 1, Tashkent 100149, Uzbekistan}

\author{Chengxun Yuan}
   \email{yuancx@hit.edu.cn }
\affiliation{School of Physics, Harbin Institute of Technology, Harbin 150001, People’s Republic of China}

\begin{abstract}
Euclidean wormholes can prepare inflation in non--Bunch--Davies initial states,
but long Lorentzian expansion screens this memory from the CMB.
We derive a visibility bound for Euclidean-matched Bogoliubov data:
the pivot excitation satisfies $|\beta_*| \lesssim e^{-2N_{\rm pre}}$,
and smooth Euclidean filters confine residual signatures to a comoving edge
$k_w=a_iM$.
Only near-minimal inflation, or an edge inside the observable window, leaves
detectable scalar, tensor, and higher-point imprints.
For longer inflation, wormhole-prepared perturbations are driven to the
Bunch--Davies prediction.
Euclidean memory therefore, becomes a quantitative bound on inflationary duration,
with direct targets in CMB polarization and large-scale structure:
the longer inflation lasts, the less of the wormhole remains on the sky.
\end{abstract}

\maketitle
Inflation converts microscopic quantum data into macroscopic correlations \cite{Guth:1980zm,Mukhanov:1981xt}. This is why any quantum-gravity account of the onset of inflation must face a precision-cosmology test: the prepared wavefunction has observational content only if its memory survives the subsequent Lorentzian expansion. A Euclidean wormhole is an imaginary-time saddle of the gravitational path integral whose analytic continuation supplies Lorentzian initial data. Recent constructions give this idea concrete form: Euclidean saddles can generate $\alpha$-like non-Bunch--Davies (non-BD) data, encoded by a negative-frequency Bogoliubov component relative to the Bunch--Davies (BD) vacuum \cite{2404.15450}, while related semiclassical weights can favor a long Lorentzian branch \cite{1904.00199,2403.17046,2411.13844,2412.03639,2603.11003}. These two ingredients pull against each other through ordinary kinematics. If the state is prepared at scale factor $a_i$, the observed CMB pivot had physical momentum $p_*=\ee^{N_{\rm pre}}H_*$ on that slice. Preserving a visible non-BD occupation at the pivot therefore requires energy of order $p_*^4$, so a few pre-pivot e-folds can turn a geometric memory into an unobservable ultraviolet burden. The duration of the Lorentzian branch, usually treated as a background detail, becomes an observational discriminator of the Euclidean origin.

De Sitter $\alpha$-vacua are unsuitable interacting low-energy endpoints: their ultraviolet and antipodal singularities generate well-known pathologies \cite{hep-th/0302050,hep-th/0306028,hep-th/0312143,hep-th/0305056,hep-th/0503022}. The controlled language is the Hadamard neighborhood of the Euclidean/BD state: smooth squeezed excitations with finite renormalized energy, adiabatic ultraviolet behavior \cite{Parker:1973qd,Birrell:1982ix,2603.00818}, and a continuation slice $a_i$ where the Euclidean kernel becomes Lorentzian Cauchy data. Trans-Planckian and initial-state studies established how such data modify inflationary spectra \cite{hep-th/0005209,hep-th/0204129,hep-th/0203198,hep-th/0209231,hep-th/0208167,hep-th/0409210}; later work developed backreaction limits \cite{0710.1302,1109.1562,1109.6566,1110.4688,1303.1440,1211.6753}, bispectrum signatures \cite{0901.4044,1104.0244,1303.1430}, and tensor or Bogoliubov-correlator probes \cite{1905.06800,2206.03667,2407.16652,2407.06258,2502.05630,2505.10534}. Euclidean matching changes the status of the initial slice. The preparation surface, ultraviolet reach, and comoving memory band are inherited from a saddle and become physical data, so observability is governed by their position relative to the CMB window.

We derive a screening bound for the pure memory carried by the matched state. The continuation surface fixes $p=k/a_i$, the Euclidean filter fixes $M$, and the comoving edge $k_w=a_iM$ marks the largest wavenumber still imprinted by the Euclidean cap. For a smooth finite-energy state, the CMB amplitude is controlled by the lever arm $N_{\rm pre}$ between preparation and pivot exit. The Supplemental Material derives the Gaussian-kernel map from Euclidean saddle data to $(\alpha_k,\beta_k)$ and the WKB origin of low-pass Euclidean profiles.

Monotone ultraviolet damping yields a pointwise bound on the pivot excitation; finite renormalized energy yields a shell-averaged bandpower bound with $\ee^{-2N_{\rm pre}}$ screening; stretched-exponential Euclidean filters add a cutoff and give $N_{\rm pre}\lesssim\ln(M/H_*)+\Order(1)$. The argument isolates inherited Euclidean memory from Lorentzian sources such as turns, particle production, or background features after matching. Those mechanisms belong to a complementary phenomenology. In the matched-state sector, BD behavior emerges from finite energy and expansion rather than from an imposed vacuum choice.

\paragraph*{Bogoliubov initial states and observable power.}
For a canonical adiabatic scalar mode during slow roll, the BD mode $u_k^{\nbd}$ is the short-distance positive-frequency solution, while a non-BD state is a squeezed admixture
\begin{equation}
 v_k(\eta)=\alpha_k u_k^{\nbd}(\eta)+\beta_k u_k^{\nbd *}(\eta),
 \qquad |\alpha_k|^2-|\beta_k|^2=1,
 \label{eq:bogodef}
\end{equation}
with $u_k^{\nbd}$ the BD mode function \cite{Bunch:1978yq,Allen:1985ux}. The coefficient $\beta_k$ is the occupation amplitude relative to BD, and $\alpha_k$ enforces canonical normalization. At late times,
\begin{equation}
 \zetaP(k)=\frac{H_*^2}{8\pi^2\epsilon_*\mpl^2}\,|\alpha_k-\beta_k|^2
 \equiv \zetaP^{\nbd}(k)\,\mathcal{F}_s(k),
 \label{eq:ps}
\end{equation}
where starred quantities are evaluated when the mode leaves the horizon and $\mathcal F_s$ is the multiplicative initial-state distortion of the scalar power. Writing $b_k\equiv|\beta_k|$, the phase-optimized deviation satisfies
\begin{equation}
 |\mathcal{F}_s-1|\le 2b_k^2+2b_k\sqrt{1+b_k^2},
 \label{eq:deltaexact}
\end{equation}
with the linear limit $|\mathcal{F}_s-1|\simeq 2b_k$ for $b_k\ll1$ \cite{1110.4688,1303.1440}. The extrema are $\mathcal{F}_{s,\max}=1+2b_k^2+2b_k\sqrt{1+b_k^2}$ and $\mathcal{F}_{s,\min}=1+2b_k^2-2b_k\sqrt{1+b_k^2}$; the phase fixes the position inside this envelope but cannot change its size.

Let the state be prepared on a Euclidean--Lorentzian matching slice $a_i$, where the imaginary-time saddle has been replaced by Lorentzian Cauchy data. The physical momentum on this slice is $p\equiv k/a_i$: for a fixed observed comoving mode, earlier preparation means larger $p$. After adiabatic subtraction, the excess renormalized energy density carried by subhorizon excitations is of order \cite{hep-th/0409210,1109.1562,1110.4688}
\begin{equation}
 \rho_{\ex}(\eta_i)\simeq\frac{1}{2\pi^2}\int_0^{\infty}\dd p\,p^3|\beta(p)|^2.
 \label{eq:rhoex}
\end{equation}
We parameterize backreaction by
\begin{equation}
 \rho_{\ex}(\eta_i)\le \xi\,\frac{\dot\phi_i^2}{2}=\xi\,\epsilon_i\mpl^2H_i^2,
 \qquad 0<\xi\lesssim1,
 \label{eq:backreaction}
\end{equation}
so $\xi$ measures the fraction of the slow-roll kinetic budget that can be stored in the excited state without destabilizing the background \cite{0710.1302,1109.1562,1303.1440}. This is the backreaction criterion for a pure initial-state effect.

\begin{figure}[t]
 \includegraphics[width=\columnwidth]{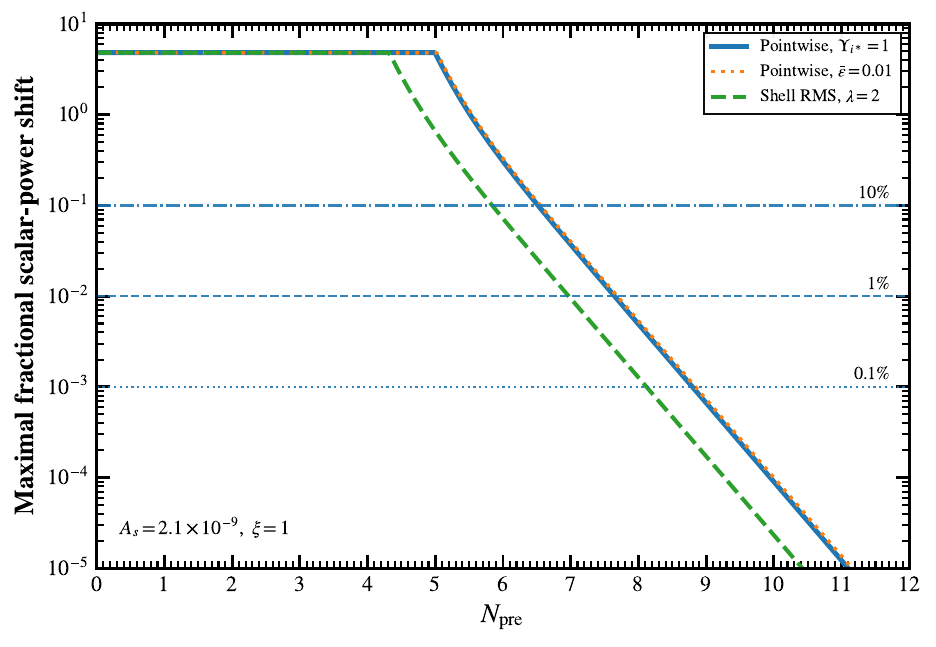}
 \caption{Screening bounds for the pivot-scale scalar-power distortion. Solid: pointwise screening bound for smooth monotone ultraviolet damping. Dashed: shell-averaged finite-energy bound with $\lambda=2$. Dotted: the drift factor $\Upsilon_{i*}$ evaluated in a constant-$\bar\epsilon=10^{-2}$ benchmark, showing that preparation-to-pivot evolution only mildly shifts the negligible-drift estimate across the observable window.}
 \label{fig:bounds}
\end{figure}

\paragraph*{Pointwise screening bound.}
Assume isotropy and smooth ultraviolet damping, so $|\beta(p)|$ is non-increasing with $p$. Equation~\eqref{eq:rhoex} then implies
\begin{equation}
 \rho_{\ex}\ge \frac{1}{2\pi^2}\int_0^p\dd q\,q^3|\beta(p)|^2
 =\frac{p^4}{8\pi^2}|\beta(p)|^2,
\end{equation}
whence
\begin{equation}
 |\beta(p)|\le \frac{\sqrt{8\pi^2\rho_{\ex}}}{p^2}.
 \label{eq:pointwisep}
\end{equation}
For the pivot mode, $p_*=k_*/a_i=\ee^{N_{\rm pre}}H_*$, where $N_{\rm pre}$ counts the e-folds between preparation and pivot exit. This is the inflationary lever arm: every extra pre-pivot e-fold raises the physical preparation momentum by $\ee$ and the energy cost by $\ee^4$. Using the observed scalar amplitude $\as\equiv H_*^2/(8\pi^2\epsilon_*\mpl^2)\simeq2.1\times10^{-9}$ \cite{1807.06211}, we obtain
\begin{equation}
 |\beta_*|\le \Upsilon_{i*}\sqrt{\frac{\xi}{\as}}\,\ee^{-2N_{\rm pre}},
 \qquad
 \Upsilon_{i*}\equiv\sqrt{\frac{\epsilon_iH_i^2}{\epsilon_*H_*^2}}.
 \label{eq:betauniversal}
\end{equation}
The factor $\Upsilon_{i*}$ tracks the preparation-to-pivot drift. In a negligible-drift quasi-de Sitter benchmark $\Upsilon_{i*}=1$, while slow roll gives $\ln\Upsilon_{i*}=\Order(\epsilon,\eta)N_{\rm pre}$. For nearly constant $\bar\epsilon$, $H_i/H_*\simeq \ee^{\bar\epsilon N_{\rm pre}}$ and $\Upsilon_{i*}\simeq \ee^{\bar\epsilon N_{\rm pre}}$; at $\bar\epsilon=10^{-2}$ and $N_{\rm pre}=8$, this gives $\Upsilon_{i*}\simeq1.08$ and moves the $1\%$ threshold by only $\Delta N_{\rm pre}\simeq0.04$. Combining Eqs.~\eqref{eq:deltaexact} and \eqref{eq:betauniversal} gives Fig.~\ref{fig:bounds}. For $\Upsilon_{i*}=1$ and $\xi=1$, ten-percent, one-percent, and permille scalar-power distortions require $N_{\rm pre}\lesssim 6.52$, $7.65$, and $8.80$. CMB-level pure memory is therefore tied to near-minimal inflation even before a wormhole filter is imposed.
\paragraph*{General finite-energy bound.}
Smooth Euclidean filters are naturally monotone in the ultraviolet. Finite energy alone gives a less local, bandpower-level statement. A narrow spike in momentum space may evade Eq.~\eqref{eq:pointwisep} at a selected momentum while keeping the integrated energy finite; observations then measure a shell average. For any fixed $\lambda>1$, define the logarithmic-shell mean
\begin{equation}
 \overline{|\beta|^2}_{\lambda}(p)
 \equiv\frac{4}{(\lambda^4-1)p^4}\int_{p}^{\lambda p}\dd q\,q^3|\beta(q)|^2.
 \label{eq:shellmean}
\end{equation}
Equation~\eqref{eq:rhoex} implies
\begin{equation}
 \overline{|\beta|^2}_{\lambda}(p)
 \le \frac{8\pi^2\rho_{\ex}}{(\lambda^4-1)p^4},
\end{equation}
so the root-mean-square excitation on every logarithmic band obeys
\begin{equation}
 \overline{|\beta|}^{\rm rms}_{\lambda,*}
 \le \Upsilon_{i*}\sqrt{\frac{\xi}{(\lambda^4-1)\as}}\,\ee^{-2N_{\rm pre}}.
 \label{eq:shellbound}
\end{equation}
Equation~\eqref{eq:shellbound} is the finite-energy bandpower statement: an isolated spike may survive, but a large \emph{coarse-grained} CMB signal cannot. If a nonmonotone profile carries an approximately constant amplitude $b_{\rm sp}$ across a narrow logarithmic shell of width $\Delta\ln p\ll 1$ centered on $p_*$, Eq.~\eqref{eq:rhoex} implies $\rho_{\ex} \gtrsim b_{\rm sp}^2 p_*^4 \Delta\ln p /(2\pi^2)$ and therefore
\begin{equation}
 \Delta\ln p \lesssim \frac{\xi\,\Upsilon_{i*}^2}{4\as \, b_{\rm sp}^2}\,\ee^{-4N_{\rm pre}}.
 \label{eq:spikewidth}
\end{equation}
A pointwise evasion therefore requires support compressed into an increasingly narrow momentum band. For smooth wormhole-motivated states, the pointwise and shell-averaged bounds converge on one conclusion: prolonged inflation screens observable memory of the preparation surface.

\begin{figure*}[t]
 \includegraphics[width=0.98\textwidth]{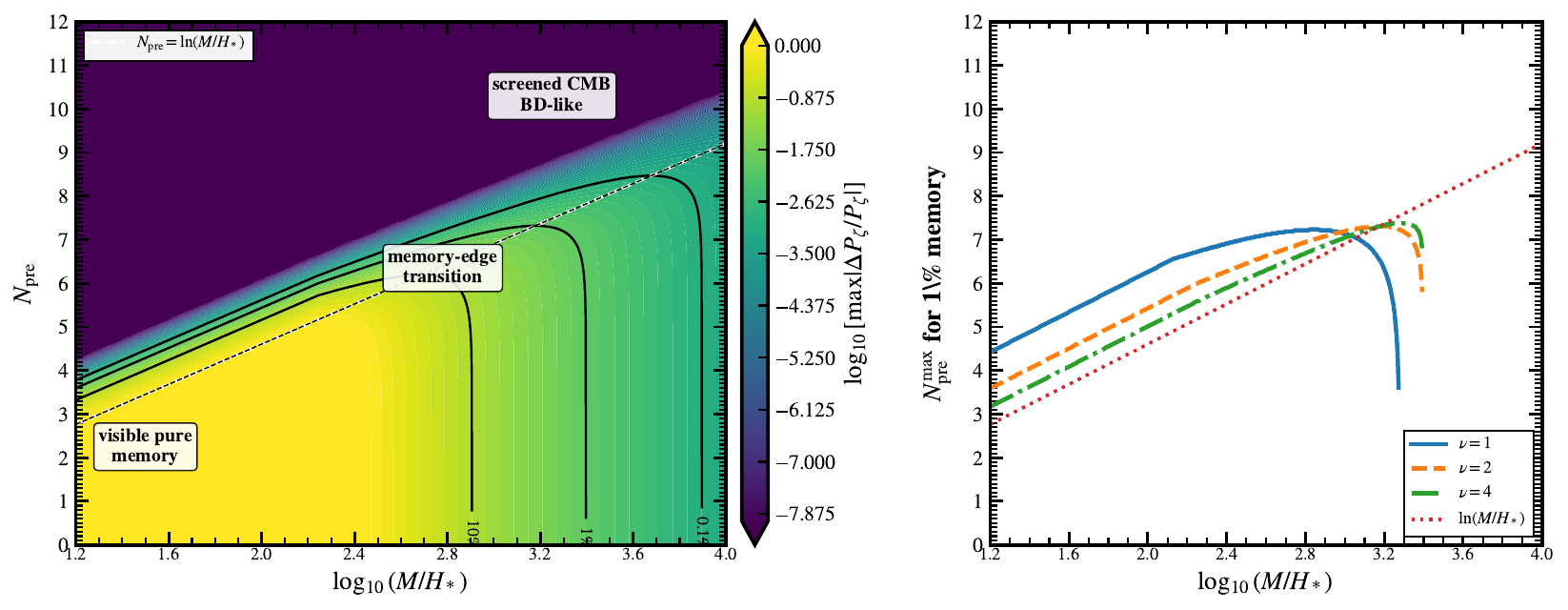}
 \caption{Visibility phase diagram for Euclidean-matched initial states. Left: maximal scalar-power memory for the Gaussian representative $\nu=2$, after saturating the backreaction bound; the contours turn wormhole initial-state phenomenology into a bound on the pre-pivot lever arm. Right: maximal $N_{\rm pre}$ compatible with a $1\%$ pivot-scale signal for several stretched-exponential shapes. The dashed curve marks $N_{\rm pre}=\ln(M/H_*)$; above it the CMB pivot lies beyond the memory edge $k_w=a_iM$ and the matched state is observationally driven toward BD.}
 \label{fig:family}
\end{figure*}

\begin{figure}[t]
 \includegraphics[width=0.96\columnwidth]{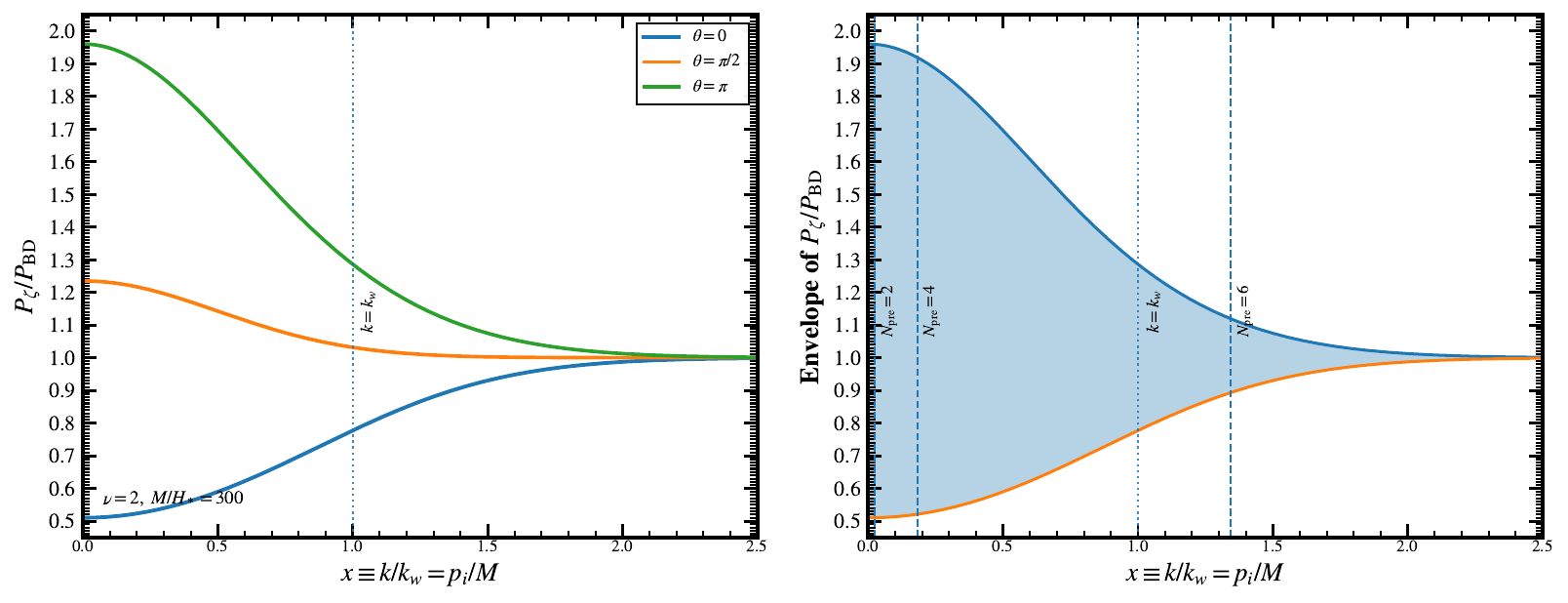}
 \caption{Left: sample spectra for the stretched-exponential filter with $\nu=2$ and $M/H_*=300$ at three phases. Right: phase envelope with three pivot locations. Smooth Euclidean preparation produces a common memory edge at $k_w=a_iM$, not a featureless high-$k$ excess.}
 \label{fig:spectra}
\end{figure}

\paragraph*{Euclidean filter family.}
Euclidean matching should not populate arbitrarily high physical momenta: a smooth saddle prepares a low-energy state with ultraviolet damping. In the Gaussian wavefunctional language detailed in the Supplemental Material, the Bogoliubov coefficient is the mismatch between the Euclidean-prepared kernel and the BD complex structure on the continuation slice; regular high-momentum Euclidean modes approach the adiabatic kernel, leaving an exponentially or super-adiabatically small mismatch \cite{Parker:1973qd,Birrell:1982ix,2603.00818}. Recent works on wormhole motivate this damping but do not uniquely fix its functional form \cite{2404.15450,2403.17046,2411.13844}. A one-parameter family captures the relevant ultraviolet behavior:
\begin{equation}
 \beta(p)=\beta_0\,\ee^{i\theta_0}\exp\!\left[-\left(\frac{p}{M}\right)^{\nu}\right],
 \qquad \nu>0,
 \label{eq:stretched}
\end{equation}
which interpolates between exponential ($\nu=1$), Gaussian ($\nu=2$), and super-Gaussian ($\nu>2$) damping. Equation~\eqref{eq:rhoex} gives
\begin{equation}
 \rho_{\ex}=\frac{M^4|\beta_0|^2}{2^{3+4/\nu}\pi^2}\,\Gamma\!\left(1+\frac{4}{\nu}\right),
 \label{eq:rhonu}
\end{equation}
so backreaction implies
\begin{equation}
 |\beta_0|\le
 \min\!\left[1,\frac{2^{2/\nu}}{\sqrt{\Gamma(1+4/\nu)}}\,\Upsilon_{i*}\sqrt{\frac{\xi}{\as}}\left(\frac{H_*}{M}\right)^2\right].
 \label{eq:beta0nu}
\end{equation}
At the pivot,
\begin{equation}
 |\beta_*|\le |\beta_0|\exp\!\left[-\left(\frac{\ee^{N_{\rm pre}}H_*}{M}\right)^{\nu}\right].
 \label{eq:betanu}
\end{equation}
Smooth Euclidean filters contain two suppressions: the backreaction factor $\ee^{-2N_{\rm pre}}$ and the filter cutoff in Eq.~\eqref{eq:betanu}. The former is the energy cost of high preparation momentum; the latter is the finite ultraviolet reach of the Euclidean cap. The dependence on the profile is mild because $2^{2/\nu}/\sqrt{\Gamma(1+4/\nu)}$ equals $1$, $\sqrt2$, and $\sqrt2$ for $\nu=1$, $2$, and $4$. The visibility phase diagram in Fig.~\ref{fig:family} displays this hierarchy. For $M/H_*=10^2$, a $1\%$ signal requires $N_{\rm pre}\lesssim5.0$--$6.3$ as $\nu$ ranges from $4$ to $1$; for $M/H_*=10^3$, the window is $N_{\rm pre}\lesssim7.1$--$7.2$. Thus
\begin{equation}
 N_{\rm pre}\lesssim \ln(M/H_*)+\Order(1),
 \label{eq:lnwindow}
\end{equation}
independent of the detailed ultraviolet profile.
\paragraph*{Memory scale, higher-point observables, and tensors.}
The screening mechanism is most transparent when expressed in terms of the comoving memory scale, the largest comoving wavenumber that still remembers the Euclidean filter,
\begin{equation}
 k_w\equiv a_iM = k_*\,\frac{M}{H_*}\,\ee^{-N_{\rm pre}}.
 \label{eq:kw}
\end{equation}
For smooth Euclidean filters, the excited band is confined to $k\lesssim k_w$. Phases change the oscillatory pattern and the sign of the modulation in Fig.~\ref{fig:spectra}, but only inside that band. Once $k_*/k_w=\ee^{N_{\rm pre}}H_*/M\gg1$, the pivot mode is exponentially driven back to the BD prediction. Phenomenology is set by the location of the memory edge, rather than a featureless high-$k$ excess.

For the low-pass Euclidean filters studied here, higher-$k$ probes do \emph{not} outrun screening: once the pivot lies outside the memory band, modes with $k>k_*$ are more strongly suppressed. The observational target is a turnover at $k\sim k_w$ shared by pure initial-state observables. The large envelope in Fig.~\ref{fig:spectra} is localized near that edge and remains compatible with the pivot-scale bound once $k_*>k_w$. Near the CMB damping tail, the leading signature is a correlated high-$\ell$ turnover in temperature and polarization rather than a broad high-$k$ excess. At shorter wavelengths, large-scale structure remains useful while the modes are linear or mildly nonlinear; spectral distortions probe dissipation of small-scale power \cite{1204.4241,1504.00675,2010.07814}; 21\,cm surveys extend the reach still further \cite{1605.09364,2501.02538}. When $k_w$ lies outside these windows, the bound predicts CMB silence rather than a residual signal.

The correlated nature of the edge is important. A scalar feature without compatible folded non-Gaussianity, tensor response, or phase coherence would point away from pure Euclidean memory and toward later Lorentzian dynamics. A turnover recurring across power spectra and phase-sensitive correlators would instead identify the continuation surface as an active part of the primordial state. The bound therefore supplies a way to read future null results and candidate anomalies in a common language: each observable measures how far its band sits from $k_w$.

The edge also governs higher-point and tensor observables. Pure initial-state corrections to folded bispectra and related higher-point functions vanish in the BD limit and require at least one explicit Bogoliubov insertion \cite{1110.4688,1104.0244,1303.1430}. For smooth filters, recent Bogoliubov-correlator and cutting-rule analyses show that each insertion carries this ultraviolet profile \cite{2407.16652,2407.06258,2502.05630}. Their hard-momentum envelope is controlled by the damping factor in Eq.~\eqref{eq:betanu}. Long inflation consequently compresses the family of pure initial-state observables toward $k\lesssim k_w$. Estimator-level bispectrum forecasts depend on survey design, whereas the damping exponent is fixed by the matched profile. The tensor sector behaves analogously,
\begin{equation}
 \tensorP(k)=\frac{2H_*^2}{\pi^2\mpl^2}|\alpha_k^{(t)}-\beta_k^{(t)}|^2,
\end{equation}
so any \emph{pure initial-state} shift of $r$ or the single-field consistency relation inherits this screening \cite{1905.06800,2206.03667}. Long inflation does not erase memory uniformly; it confines primordial memory toward the infrared scale $k_w$.
\paragraph*{Discussion.}
Euclidean-wormhole initial data and a long Lorentzian branch belong to one semiclassical narrative \cite{2404.15450,1904.00199,2403.17046,2411.13844,2412.03639,2603.11003}. The screening bound turns that narrative into the ratio $k_*/k_w=\ee^{N_{\rm pre}}H_*/M$. When the saddle prepares a smooth Hadamard matched state, the CMB pivot carries Euclidean memory only inside $k\lesssim k_w$; moving the preparation surface deeper into the past pushes the pivot through the Euclidean cutoff and into the BD basin. Correlator technology for Bogoliubov states \cite{2407.16652,2407.06258,2502.05630} and current CMB analyses of non-BD departures \cite{2505.10534} then acquire a physical scale fixed by the preparation geometry.

With this bound in place, any detected non-BD feature gains diagnostic content. A CMB-scale signal would select a short Euclidean-to-pivot lever arm, a momentum profile concentrated near the pivot, or new Lorentzian dynamics after the matching surface. The shell-averaged bound quantifies the price of localization through Eq.~\eqref{eq:spikewidth}; later dynamics can imprint structure as an independent source, as in controlled deformations of the inflationary evolution \cite{2603.00818}. Conversely, a smooth spectrum across the measured window constrains $\ee^{N_{\rm pre}}H_*/M$ and places the Euclidean edge beyond the accessible band. The visibility map locates where wormhole memory can reside, beyond the loose statement that excited states redshift away.

High-$\ell$ CMB temperature and polarization test an edge near the damping tail; large-scale structure follows the turnover while the modes remain perturbative. Spectral distortions probe the dissipation of smaller-scale scalar power \cite{1204.4241,1504.00675,2010.07814}, and 21\,cm observations extend the lever arm still further \cite{1605.09364,2501.02538}. Folded bispectra, tensor consistency tests, and Bogoliubov cutting relations inherit the infrared boundary because their pure initial-state terms contain explicit powers of $\beta_k$. Swampland-distance conjectures may reduce the effective cutoff \cite{1812.07558}; the screening bound uses only finite energy and matched-state data.

Long inflation acts as a dynamical filter on quantum-gravitational initial data. It does not erase the preparation surface indiscriminately; it moves its observable imprint to the comoving edge $k_w$. Inside that edge, a Euclidean cap can leave correlated scalar, tensor, and higher-point structure. Beyond it, smooth wormhole-prepared states become BD-like for reasons intrinsic to the low-energy state: finite energy, ultraviolet regularity, and the exponential lever arm of inflation. The duration of inflation is thereby promoted from a background parameter to a discriminator of primordial origin: it decides whether Euclidean preparation is visible, displaced to smaller scales, or observationally folded into the BD vacuum. Screening thereby converts quantum-gravitational boundary data into a measurable scale, $k_w$.
\bibliography{refs}
\setcounter{equation}{0}
\setcounter{figure}{0}
\setcounter{table}{0}
\onecolumngrid
\renewcommand{\theequation}{S\arabic{equation}}
\renewcommand{\thefigure}{S\arabic{figure}}
\renewcommand{\thetable}{S\arabic{table}}
\renewcommand{\theHequation}{S\arabic{equation}}
\renewcommand{\theHfigure}{S\arabic{figure}}
\renewcommand{\theHtable}{S\arabic{table}}
\renewcommand{\thesection}{S\Roman{section}}

\section*{Supplemental Material: Long Inflation Screens Euclidean-Wormhole Initial States}

This Supplemental Material gives the Euclidean matching construction, the finite-energy screening bounds used in the Letter, and the formulas entering the figures.

\section{Euclidean matching as a Gaussian kernel}
The main text treats the Euclidean saddle through the low-energy Lorentzian state it prepares. That matched state contains the information relevant for the visibility bound. For each canonically normalized scalar perturbation mode $q_k$ with conjugate momentum $\pi_k$, a regular semiclassical saddle fixes a Gaussian wavefunctional on the continuation slice,
\begin{equation}
\Psi_i[q]\propto \exp\!\left[-\frac12\int\frac{\dd^3k}{(2\pi)^3}\,\Omega_k^{E}\,q_kq_{-k}\right],
\qquad {\rm Re}\,\Omega_k^{E}>0 .
\label{eqS:gaussian_kernel}
\end{equation}
Here $\Omega_k^{E}$ is the Euclidean-prepared kernel after analytic continuation to the Lorentzian Cauchy surface. The BD state on the continuation slice defines a reference complex structure with kernel $\Omega_k^{\rm BD}$. A pure Gaussian state with kernel $\Omega_k^{E}$ is equivalently a Bogoliubov transform of that BD state. With the standard canonical normalization of the pair $(q_k,\pi_k)$,
\begin{equation}
 |\beta_k|^2=\frac{|\Omega_k^{E}-\Omega_k^{\rm BD}|^2}{4\,{\rm Re}\,\Omega_k^{E}\,{\rm Re}\,\Omega_k^{\rm BD}},
\label{eqS:kernel_beta}
\end{equation}
up to the harmless phase convention used to define $\alpha_k$ and $\beta_k$. Thus the observable excitation is the mismatch between the Euclidean kernel and the BD complex structure on the matching surface; after the saddle and continuation slice are fixed, $\beta_k$ is state data rather than a tunable regulator.

The Gaussian representation also clarifies the low-pass character of smooth Euclidean preparation. At large physical momentum $p=k/a_i$, locality and adiabatic regularity make the Euclidean mode equation take the WKB form
\begin{equation}
\left[-\partial_\tau^2+\omega_E^2(\tau,p)\right]f_p(\tau)=0,
\qquad \omega_E(\tau,p)=p\left[1+\Order\!\left(\frac{H_i^2}{p^2}\right)\right].
\label{eqS:wkb_mode}
\end{equation}
The regular Euclidean solution is the decaying branch in the cap. Its Lorentzian continuation gives
\begin{equation}
\Omega_p^{E}=\Omega_p^{\rm ad}+\delta\Omega_p,
\qquad
\frac{|\delta\Omega_p|}{p}\lesssim C(p)\exp[-2S_E(p)],
\label{eqS:kernel_wkb}
\end{equation}
where $\Omega_p^{\rm ad}$ is the adiabatic kernel \cite{Parker:1973qd,Birrell:1982ix}, $C(p)$ is at most a slow function of $p$ in the controlled regime, and
\begin{equation}
S_E(p)=\int_{\cal C}\dd\tau\,\omega_E(\tau,p)
\label{eqS:SE}
\end{equation}
measures the Euclidean damping accumulated before the continuation surface. Combining Eqs.~\eqref{eqS:kernel_beta} and \eqref{eqS:kernel_wkb} gives
\begin{equation}
|\beta(p)|\lesssim B(p)\exp[-2S_E(p)] + \Order\!\left(\frac{H_i^2}{p^2}\right)_{\rm ad.\,sub.}.
\label{eqS:generic_filter}
\end{equation}
Adiabatic subtraction removes the local power-series part; the remaining nonadiabatic memory is suppressed by the Euclidean action. A saddle with ultraviolet scale $M$ produces $|\beta(p)|\sim B(p)\exp[-(p/M)^\nu]$ after order-unity constants are absorbed into $M$. The exponent $\nu$ depends on the analytic structure of the cap and on the effective barrier; the visibility bound in the Letter depends only on finite energy and the resulting memory edge.

In an arbitrary Bogoliubov parameterization, $a_i$ and $\beta(p)$ can be moved as regulator-like data. Euclidean matching gives them physical content: $a_i$ is the continuation surface and $M$ is the ultraviolet range over which the saddle can imprint the Lorentzian state. The invariant ratio is
\begin{equation}
\frac{k_*}{k_w}=\frac{\ee^{N_{\rm pre}}H_*}{M},
\qquad k_w\equiv a_iM.
\label{eqS:edge_ratio}
\end{equation}
The Letter's criterion is a bound on this ratio. Smooth changes in the saddle modify $B(p)$, $M$, or $\nu$; the finite-energy scaling with $p_*$ and the memory edge $k_w$ remain.

\section{Phase envelope}
Write $\beta_k=b_k\ee^{i\theta_k}$ and $\alpha_k=\sqrt{1+b_k^2}\,\ee^{i\varphi_k}$; for the envelope this is an optimization parameterization, not an extra dynamical assumption. Then
\begin{equation}
\mathcal{F}_s(k)=|\alpha_k-\beta_k|^2=1+2b_k^2-2b_k\sqrt{1+b_k^2}\cos(\varphi_k-\theta_k).
\end{equation}
Maximizing over the relative phase gives
\begin{equation}
\mathcal{F}_{s,\max}=1+2b_k^2+2b_k\sqrt{1+b_k^2},\qquad
\mathcal{F}_{s,\min}=1+2b_k^2-2b_k\sqrt{1+b_k^2},
\end{equation}
so the envelope bound quoted in the Letter follows. For $b_k\ll1$, $\mathcal{F}_{s,\max}-1=2b_k+2b_k^2+\Order(b_k^3)$ and $1-\mathcal{F}_{s,\min}=2b_k-2b_k^2+\Order(b_k^3)$.

For a desired upper-envelope fractional distortion $\delta\equiv \mathcal F_{s,\max}-1$, the inverse relation is
\begin{equation}
 b_{\delta}=\frac{\delta}{2\sqrt{1+\delta}},
\label{eqS:bdeltainv}
\end{equation}
which reduces to $b_\delta\simeq\delta/2$ for small distortions. In the negligible-drift benchmark, the pointwise bound gives
\begin{equation}
N_{\rm pre}^{\rm max}(\delta)=\frac12\ln\!\left[\frac{\sqrt{\xi/A_s}}{b_\delta}\right]
\label{eqS:Nmaxdelta}
\end{equation}
before imposing an explicit Euclidean filter. This expression gives $N_{\rm pre}^{\rm max}=6.52$, $7.65$, and $8.80$ for $\delta=0.1$, $0.01$, and $10^{-3}$ at $\xi=1$. The numerical thresholds quoted in the Letter follow from the phase envelope rather than the linearized approximation.

\section{Universal screening bounds}
At the preparation slice $a_i$, the physical momentum is $p=k/a_i$ and the renormalized excitation energy is estimated by
\begin{equation}
\rho_{\ex}(\eta_i)\simeq \frac{1}{2\pi^2}\int_0^{\infty}\dd p\,p^3|\beta(p)|^2,
\qquad
\rho_{\ex}(\eta_i)\le \xi\,\epsilon_i\mpl^2H_i^2.
\label{eqS:rho_back}
\end{equation}
Under isotropic monotone ultraviolet damping,
\begin{equation}
\rho_{\ex}\ge \frac{1}{2\pi^2}\int_0^p\dd q\,q^3|\beta(p)|^2=\frac{p^4}{8\pi^2}|\beta(p)|^2,
\end{equation}
so $|\beta(p)|\le \sqrt{8\pi^2\rho_{\ex}}/p^2$. At the pivot, $p_*=\ee^{N_{\rm pre}}H_*$ and , and the observed scalar amplitude is
\begin{equation}
\as=\frac{H_*^2}{8\pi^2\epsilon_*\mpl^2}.
\end{equation}
Combining this with Eq.~\eqref{eqS:rho_back} gives
\begin{equation}
|\beta_*|\le \Upsilon_{i*}\sqrt{\frac{\xi}{\as}}\,\ee^{-2N_{\rm pre}},
\qquad
\Upsilon_{i*}\equiv\sqrt{\frac{\epsilon_iH_i^2}{\epsilon_*H_*^2}}.
\label{eqS:betauniv}
\end{equation}
If $\epsilon$ is nearly constant, $\Upsilon_{i*}\simeq \ee^{\bar\epsilon N_{\rm pre}}$; for $\bar\epsilon=10^{-2}$ and $N_{\rm pre}=8$, this gives $\Upsilon_{i*}\simeq1.08$ and $\Delta N_{\rm pre}\simeq \tfrac12\ln\Upsilon_{i*}\simeq0.04$.

Finite energy alone gives the shell-averaged statement. Define
\begin{equation}
\overline{|\beta|^2}_{\lambda}(p)\equiv \frac{4}{(\lambda^4-1)p^4}\int_p^{\lambda p}\dd q\,q^3|\beta(q)|^2,
\qquad \lambda>1.
\label{eqS:shell}
\end{equation}
By construction $\overline{|\beta|^2}_{\lambda}(p)\to |\beta(p)|^2$ as $\lambda\to1^+$. Using Eq.~\eqref{eqS:rho_back},
\begin{equation}
\overline{|\beta|}^{\rm rms}_{\lambda,*}\le \Upsilon_{i*}\sqrt{\frac{\xi}{(\lambda^4-1)\as}}\,\ee^{-2N_{\rm pre}}.
\label{eqS:shellbound}
\end{equation}
A spike of amplitude $b_{\rm sp}$ spread over a narrow shell $\Delta\ln p\ll1$ around $p_*$ instead obeys
\begin{equation}
\Delta\ln p\lesssim \frac{\xi\,\Upsilon_{i*}^2}{4\as\,b_{\rm sp}^2}\,\ee^{-4N_{\rm pre}},
\end{equation}
so a pointwise evasion requires exponentially narrow support.

\section{Stretched-exponential filters and the memory scale}
A flexible representation of smooth Euclidean filters is
\begin{equation}
\beta(p)=\beta_0\,\ee^{i\theta_0}\exp\!\left[-\left(\frac{p}{M}\right)^{\nu}\right],\qquad \nu>0.
\label{eqS:nuansatz}
\end{equation}
Then
\begin{equation}
\rho_{\ex}=\frac{M^4|\beta_0|^2}{2^{3+4/\nu}\pi^2}\,\Gamma\!\left(1+\frac{4}{\nu}\right),
\end{equation}
which implies
\begin{equation}
|\beta_0|\le \min\!\left[1,\frac{2^{2/\nu}}{\sqrt{\Gamma(1+4/\nu)}}\,\Upsilon_{i*}\sqrt{\frac{\xi}{\as}}\left(\frac{H_*}{M}\right)^2\right].
\label{eqS:beta0}
\end{equation}
At the pivot,
\begin{equation}
|\beta_*|\le |\beta_0|\exp\!\left[-\left(\frac{\ee^{N_{\rm pre}}H_*}{M}\right)^{\nu}\right].
\label{eqS:betanu}
\end{equation}
For $\nu=1,2,4$ the normalization factor in Eq.~\eqref{eqS:beta0} is order unity, so the visibility window is governed mainly by $N_{\rm pre}\lesssim\ln(M/H_*)+\Order(1)$. The associated comoving memory scale is
\begin{equation}
k_w\equiv a_iM = k_*\,\frac{M}{H_*}\,\ee^{-N_{\rm pre}}.
\end{equation}
The large envelope is localized near $k\sim k_w$; if $k_*>k_w$, the pivot remains screened.

\section{Higher-point and tensor remarks}
Pure initial-state bispectra and related higher-point terms vanish when $\beta\to0$ and therefore require at least one explicit Bogoliubov insertion \cite{1110.4688,1104.0244,1303.1430}. For smooth filters, recent Bogoliubov-correlator and cutting-rule analyses show that the corresponding ultraviolet profile governs each insertion \cite{2407.16652,2407.06258,2502.05630}, so the hard-momentum envelope is again controlled by $k_{\rm hard}/k_w$. For tensors,
\begin{equation}
\tensorP(k)=\frac{2H_*^2}{\pi^2\mpl^2}\,|\alpha_k^{(t)}-\beta_k^{(t)}|^2,
\end{equation}
so a larger effect would trace later dynamics outside the matched-state analysis.

\section{Phenomenological domain and relation to earlier bounds}
The screening bound uses the familiar physical ingredients that make excited inflationary states controllable: a Hadamard ultraviolet structure, finite renormalized energy, and small backreaction. Euclidean matching changes their physical meaning. Earlier bounds constrain how large an arbitrary excited state may be on an adopted initial slice. Here the slice and the ultraviolet memory scale have physical meaning because they are supplied by the Euclidean--Lorentzian matching. The CMB pivot either lies inside the comoving memory band $k\lesssim k_w$ or has already crossed beyond it.

The pointwise result requires smooth monotone ultraviolet damping. Dropping monotonicity leaves the shell-averaged bound, so a large feature can survive only by occupying an exponentially narrow logarithmic interval. Without smoothness or finite energy the construction leaves the controlled matched-state class and reintroduces the ultraviolet pathologies that motivate working near the Euclidean/BD vacuum. Later nonadiabatic dynamics, turns in field space, particle production, or features in the inflationary background can repopulate observable modes; those are new Lorentzian sources rather than pure memory of the Euclidean preparation surface.

A CMB-scale non-BD detection in this framework would point to a short Euclidean-to-pivot lever arm, to a highly localized nonmonotone profile, or to later dynamics after the matching surface. For smooth matched states, the absence of such a feature bounds the ratio $k_*/k_w$ and therefore the combination $\ee^{N_{\rm pre}}H_*/M$. Long inflation thereby converts the lever arm into an observational censor of Euclidean-wormhole initial data.

\end{document}